\documentclass[twocolumn,aps,amsmath,amssymb,floatfix,superscriptaddress,prb,longbibliography]{revtex4-1}
\usepackage{graphicx}
\usepackage{dcolumn}
\usepackage{bm}
\usepackage{xspace}
\usepackage{hyperref}
\usepackage{color}
\usepackage[flushleft]{threeparttable}

\def\ncso{Na$_3$Co$_2$SbO$_6$\xspace}
\def\nczso{Na$_3$Co$_{2-x}$Zn$_x$SbO$_6$\xspace}
\begin{document}

\title{Signatures of a gapless quantum spin liquid in the Kitaev material \nczso}

\author{Zhongtuo~Fu}
\author{Ruokai~Xu}
\author{Yingqi~Chen}
\affiliation{Institute for Advanced Materials, Hubei Normal University, Huangshi 435002, China}
\author{Song~Bao}
\affiliation{National Laboratory of Solid State Microstructures and Department of Physics, Nanjing University, Nanjing 210093, China}
\author{Hong~Du}
\affiliation{Tsung-Dao Lee Institute $\&$ School of Physics and Astronomy, Shanghai Jiao Tong University, Shanghai 200240, China}
\author{Jiahua~Min}
\author{Shuhan~Zheng}
\author{Yongjun~Zhang}
\author{Meifeng~Liu}
\author{Xiuzhang~Wang}
\author{Hong~Li}
\affiliation{Institute for Advanced Materials, Hubei Normal University, Huangshi 435002, China}
\author{Ruidan~Zhong}
\email{rzhong@sjtu.edu.cn}
\affiliation{Tsung-Dao Lee Institute $\&$ School of Physics and Astronomy, Shanghai Jiao Tong University, Shanghai 200240, China}
\author{Huiqian~Luo}
\affiliation{Beijing National Laboratory for Condensed Matter Physics, Institute of Physics, Chinese Academy of Sciences, Beijing 100190, China}
\affiliation{Songshan Lake Materials Laboratory, Dongguan, Guangdong 523808, China}
\author{Jun-Ming~Liu}
\affiliation{National Laboratory of Solid State Microstructures and Department of Physics, Nanjing University, Nanjing 210093, China}
\affiliation{Collaborative Innovation Center of Advanced Microstructures, Nanjing University, Nanjing 210093, China}
\author{Zhen~Ma}
\email{zma@hbnu.edu.cn}
\affiliation{Institute for Advanced Materials, Hubei Normal University, Huangshi 435002, China}
\affiliation{State Key Laboratory of Surface Physics and Department of Physics, Fudan University, Shanghai 200433, China}
\author{Jinsheng~Wen}
\email{jwen@nju.edu.cn}
\affiliation{National Laboratory of Solid State Microstructures and Department of Physics, Nanjing University, Nanjing 210093, China}
\affiliation{Collaborative Innovation Center of Advanced Microstructures, Nanjing University, Nanjing 210093, China}


\begin{abstract}
The honeycomb-lattice cobaltate \ncso has recently been proposed to be a proximate Kitaev quantum spin liquid~(QSL) candidate. However, non-Kitaev terms in the Hamiltonian lead to a zigzag-type antiferromagnetic~(AFM) order at low temperatures. Here, we partially substitute magnetic Co$^{2+}$ with nonmagnetic Zn$^{2+}$ and investigate the chemical doping effect in tuning the magnetic ground states of \nczso. X-ray diffraction characterizations reveal no structural transition but quite tiny changes on the lattice parameters over our substitution range $0\leq x\leq0.4$. Magnetic susceptibility and specific heat results both show that AFM transition temperature is continuously suppressed with increasing Zn content $x$ and neither long-range magnetic order nor spin freezing is observed when $x\geq0.2$. More importantly, a linear term of the specific heat representing fermionic excitations is captured below 5~K in the magnetically disordered regime, as opposed to the $C_{\rm m}\propto T^3$ behavior expected for bosonic excitations in the AFM state. Based on the data above, we establish a magnetic phase diagram of \nczso. Our results indicate the presence of gapless fractional excitations in the samples with no magnetic order, evidencing a potential QSL state induced by doping in a Kitaev system.
\end{abstract}

\maketitle

\section{Introduction}

As an alternative avenue to the long-sought quantum spin liquid (QSL) state, the exactly solvable Kitaev honeycomb model has motivated extensive research interest in recent years~\cite{RevModPhys.89.025003,Broholmeaay0668,nrp1_264,TREBST20221,0953-8984-29-49-493002}. It features bond-dependent Ising interactions between effective spin-1/2 nearest neighbors on the honeycomb lattice that lead to an exchange frustration of the spin on a single site~\cite{aop321_2}. The intrinsic frustrations will introduce strong quantum fluctuations, and thus give rise to a magnetically disordered state~\cite{aop321_2,nrp1_264,TREBST20221,0953-8984-29-49-493002}, which is so called Kitaev QSLs~\cite{aop321_2}. Such a state hosts long-range quantum entanglement and fractional excitations represented by Majorana fermions that can find potential applications in fault-tolerant quantum computation~\cite{Kitaev20032,aop321_2,RevModPhys.80.1083,Barkeshli722}. Therefore, materializing the Kitaev model has currently become a topical issue in the field of strongly correlated electrons.

Following the seminal work proposed by Jackeli and Khaliullin in 2009~\cite{prl102_017205}, the Kitaev QSL state has been sought after in a Mott insulator with strong spin-orbit coupling~(SOC). It subsequently stimulates extensive studies in real materials with heavy $d^5$ transition metal ions~\cite{0953-8984-29-49-493002,nrp1_264,doi:10.1021/acs.chemrev.0c00641,TREBST20221}, including 5$d^5$ $A_2$IrO$_3$ ($A$ = Na, Li, and Cu) family and 4$d^5$ $\alpha$-RuCl$_3$. Although Na$_2$IrO$_3$ as a representative member of 5$d$ iridates is the first proposed Kitaev candidate material~\cite{prl105_027204}, the unavailability of large-size and high-quality single crystals~\cite{0953-8984-29-49-493002} restricts in-depth studies. By contrast, more efforts are devoted into 4$d$ ruthenates. Now it is well established that $\alpha$-RuCl$_3$ harbors a zigzag magnetic ground state~\cite{PhysRevB.91.144420,PhysRevB.92.235119,nm15_733}, but dominant Kitaev interactions are unambiguously unveiled~\cite{PhysRevLett.118.107203,nm15_733,Banerjee1055,np13_1079}. Further researches point out that its zigzag order is fragile and can be efficiently suppressed by a moderate magnetic field applied within the honeycomb plane~\cite{PhysRevB.99.140413,PhysRevB.102.140402,nc10_2470,PhysRevLett.120.077203,PhysRevResearch.2.013014,nc12_4007,Xiaoxue_Zhao:57501}. Last but not least, the quantized thermal Hall conductivity at an in-plane field is observed and it is interpreted as the evidence of itinerant Majorana fermions expected in a Kitaev QSL~\cite{nature559_227,doi:10.1126/science.aay5551}. However, there are subsequent reports challenging the claim and proposing that the thermal Hall effect is actually predominantly resulting from  phonons~\cite{PhysRevX.12.021025,PhysRevB.106.L041102}. The contradiction of various results leaves the nature of the field-driven magnetically disordered phase an open question~\cite{PhysRevLett.120.067202,PhysRevB.103.054440,PhysRevLett.120.117204,nature559_227,PhysRevLett.120.217205,np17_915,doi:10.1126/science.aay5551,np18_401,nm21_416,PhysRevLett.125.037202}. Besides that, the division of magnetic phase diagram is also a hotly debated issue~\cite{PhysRevLett.119.227208,PhysRevB.95.180411,PhysRevLett.119.037201,npjqm3_8,PhysRevB.101.020414}. Considering the limited choices of real Kitaev materials as well as the remaining controversies, the search for Kitaev QSLs in a wider range of quantum magnets is thus highly desirable.

A more recent proposal emphasizes that $d^7$ Co$^{2+}$ with a high-spin $t^5_{2g}e^2_g$ configuration may serve as another promising platform to study Kitaev physics~\cite{PhysRevB.97.014407,PhysRevB.97.014408,PhysRevLett.125.047201}. Such Co$^{2+}$ ions under an octahedral crystal field of oxygens allow a Kramers doublet ground state with effective spin $J_{\rm eff}$ = 1/2 that can realize Kitaev interactions, despite relatively weak SOC effect compared with its 4$d$/5$d$ counterparts. Along this line, a 3$d^7$ honeycomb-layered cobaltate \ncso is proposed to be proximate to the Kitaev QSL phase~\cite{PhysRevB.106.014413,PhysRevLett.125.047201,PhysRevB.106.174411,PhysRevX.12.041024}. Neutron diffraction experiments have revealed the zigzag configuration of moments in the $ab$ plane~\cite{WONG201618,PhysRevMaterials.3.074405}, as the cases of $d^5$ iridates and ruthenates~\cite{npjqm4_12,doi:10.1021/acs.chemrev.0c00641}. Accumulating inelastic neutron scattering~(INS) results confirm a spin-orbit assisted $J_{\rm eff}$ = 1/2 ground state characterized by the spin-orbit excitons distributed between 20 and 30~meV~\cite{PhysRevB.102.224429,Kim_2021}, and low-energy spin dynamics can be well described by a Heisenberg-Kitaev model~\cite{PhysRevB.102.224429,Kim_2021,Kim2021,doi:10.1142/S0217979221300061,PhysRevB.106.014413}. Highly anisotropic spin interactions in the $ab$ plane are also revealed by magnetometry on twin-free single crystals~\cite{PhysRevX.12.041024}. Further study with applying magnetic fields brings a gapped spin-liquid-like phase around 2~T~\cite{PhysRevB.107.054411}. These results raise a question: Is it possible to find the QSL state under zero magnetic field by doping \ncso? In this paper, we investigate the Zn-doping effect in tuning the magnetic ground states of \nczso. It is found that a gapless QSL phase may appear for the $x\geq0.2$ regime.

\section{Experimental Details}

Polycrystalline samples of \nczso with $0\leq x\leq0.4$ and nonmagnetic Na$_3$Zn$_2$SbO$_6$ were synthesized by solid-state reaction method. The raw materials Co$_3$O$_4$~(99.99\%), ZnO~(99.99\%), and Sb$_2$O$_3$~(99.99\%) were weighed with a stoichiometric ratio, whereas 10\% excess Na$_2$CO$_3$~(99.99\%) was added to compensate for the loss due to volatilization upon heating. The precursor powders were mixed and thoroughly ground in an agate mortar. Then they were loaded into the alumina crucibles with a lid and preheated at 800~$^\circ$C in air for 30 h. After that, they were reground and sintered at 900~$^\circ$C for 60 h with one intermediate grinding.

X-ray diffraction~(XRD) data were collected at room temperature in an X-ray diffractometer (SmartLab SE, Rigaku) using the Cu-$K_\alpha$ edge with a wavelength of 1.54~\AA. In the measurements, the scan range of 2$\theta$ is from 10$^\circ$ to 80$^\circ$ with a step of 0.02$^\circ$ and a rate of 10$^\circ$/min, respectively. Rietveld refinements on XRD data were performed by GSAS software. Dc magnetic susceptibility was measured on powder samples with a typical mass $\sim$10~mg. This procedure was performed within 2-300~K in a Quantum Design physical property measurement system~(PPMS, Dynacool), equipped with a vibrating sample magnetometer~(VSM) option. Ac magnetic susceptibility measurements were performed on pelletized samples with a typical mass $\sim$5~mg within 2-100~K using an ac susceptibility measurement option in a PPMS Dynacool, in which an ac field of 0.2 Oe was applied. Specific heat was measured within 2-20~K on pelletized samples with a typical mass $\sim$3~mg in a PPMS Dynacool.

\section{Results}
\subsection{Crystal structure and X-ray diffraction}

\begin{figure*}[htb]
\centerline{\includegraphics[width=6.8in]{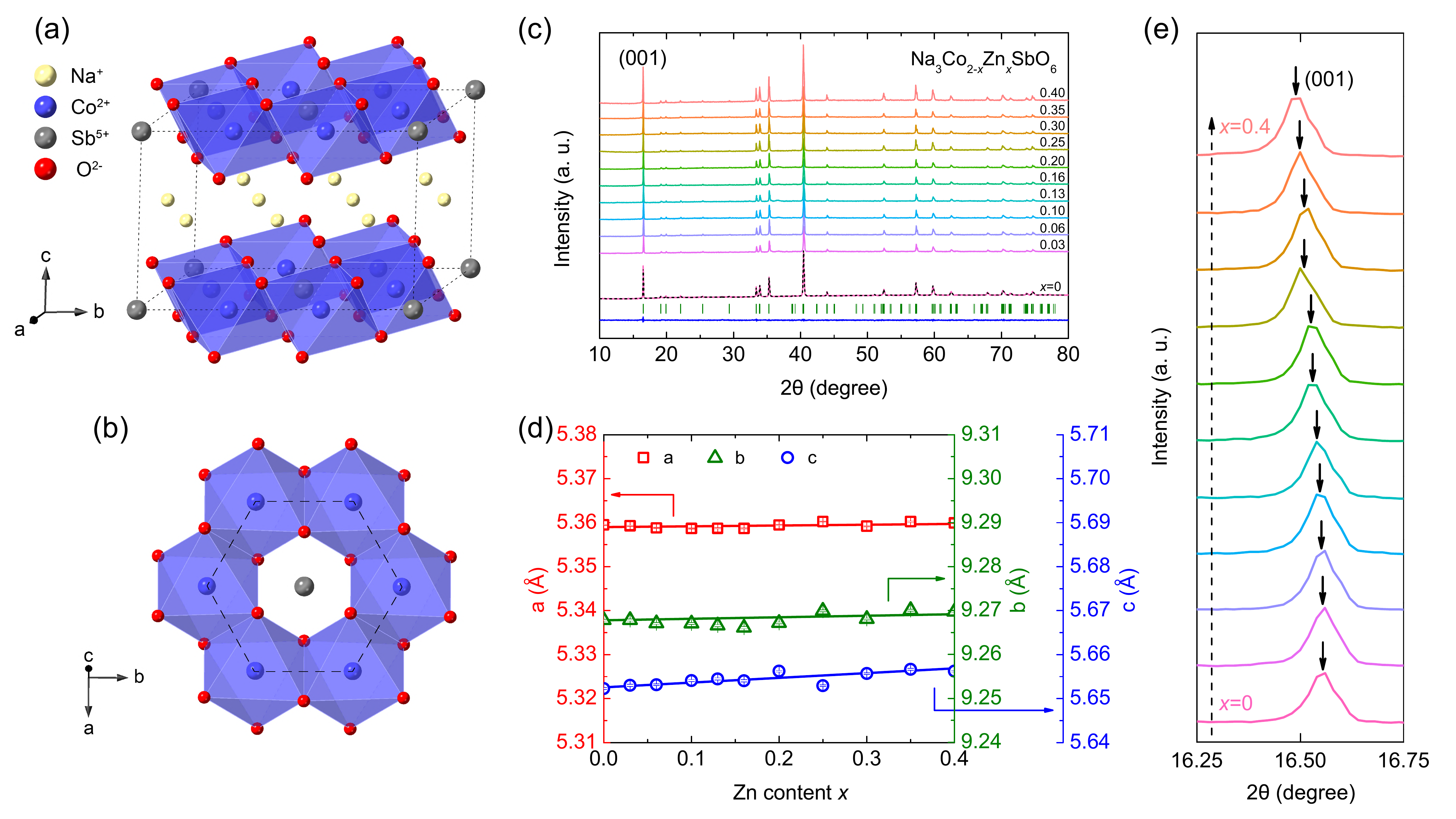}}
\caption{
(a) Schematic crystal structure of \ncso. (b) Top view of the honeycomb layer of edge-shared CoO$_6$ octahedra. Dashed lines in (a) and (b) represent the crystalline unit cell and the honeycomb unit formed by magnetic Co$^{2+}$ ions, respectively. (c) XRD patterns of the compound series \nczso with $0\leq x\leq0.4$. Green ticks denote Bragg peak positions of \ncso with space group $C2/m$ (No. 12). The black dashed line represents the calculated results for the $x$ = 0 compound, and the bottom blue line represents the difference between the experiments and the calculations for this compound. (d) Evolution of the lattice parameters $a$, $b$, and $c$ with varying the zinc concentration $x$. The solid lines are guides to the eye. (e) Zoom-in view of a typical diffraction peak (001) for different $x$. Arrows indicate the positions of the (001) peak. Errors represent one standard deviation throughout the paper.
\label{fig1}}
\end{figure*}

Figures~\ref{fig1}(a) and \ref{fig1}(b) show the schematics of the monoclinic crystal structure and a two-dimensional honeycomb layer of \ncso, respectively. This compound has the same space group $C2/m$ as $\alpha$-RuCl$_3$~\cite{PhysRevB.92.235119,PhysRevB.93.134423}. The magnetic honeycomb layers consist of six edge-shared CoO$_6$ octahedra with one SbO$_6$ octahedron located at the center of each hexagon. They arrange along the $c$ axis and are separated by an intermediate nonmagnetic layer of Na$^+$ ions. Since the nonmagnetic layers block Co-Co superexchange interaction pathways along the $c$ axis, magnetic couplings are defined within the $ab$ plane, featuring a quasi-two-dimensional magnetism~\cite{PhysRevMaterials.3.074405}.

In Fig.~\ref{fig1}(c), we present the XRD patterns of \nczso as well as the Rietveld refinement results of the parent compound. The refinements were performed based on a perfect structure model, which was used to process XRD data of polycrystal samples in the previous works (Refs.~\onlinecite{VICIU20071060,WONG201618}). For the $x$ = 0 compound, it gives rise to the lattice parameters $a$ = 5.3595(1)~\AA, $b$ = 9.2678(2)~\AA, $c$ = 5.6523(2)~\AA, and $\beta =$ 108.482(1)$^\circ$, which are comparable to the existing literatures~\cite{VICIU20071060,WONG201618,PhysRevMaterials.3.074405}. The refinement parameters are $R_{\rm p} \approx$ 2.80\%, $R_{\rm wp} \approx$ 3.88\%, and $\chi^2 \approx$ 1.726, respectively, suggesting a high purity of our synthesized samples. When increasing the zinc content $x$, no peak splitting or additional reflections suggestive of a structural transition are observed in the whole compound series. Thus, we start with the original structure model of \ncso~\cite{VICIU20071060,WONG201618} and perform the refinements for $x$ $>$ 0 compounds. It is found that all the measured reflections can be well indexed with the space group $C2/m$. The extracted lattice parameters are depicted in Fig.~\ref{fig1}(d). Parameters $a$ and $b$ remain almost constant whereas $c$ slightly rises with increasing $x$. In other words, the unit cell is a bit stretched along the $c$ axis, indicating an enhancement of the quasi-two-dimensional property with Zn being introduced into the system. Nevertheless, the overall changes are less than 0.1\% over the whole substitution range of $0\leq x\leq0.4$. Figure~\ref{fig1}(e) shows the zoom-in view of a typical diffraction peak (001) for different $x$. There is indeed no splitting but a tiny shift to the low-angle direction with increasing $x$, and the shift range is just within 0.06$^\circ$. It implies little change on the crystal structure as depicted in Fig.~\ref{fig1}(d). It is reasonable since the ion radii of Zn$^{2+}~(\sim$0.74~\AA) and of high-spin Co$^{2+}~(\sim$0.745~\AA) for octahedral coordination are quite close~\cite{https://doi.org/10.1111/j.1151-2916.2003.tb03575.x}, and thus a low Zn-doped level within 20\% leads to a negligible change on the crystal structure. Analogous cases are also observed in Zn-doped cobaltates Zn$_{1-x}$Co$_x$TiO$_3$ and BaCo$_{1-x}$Zn$_x$SiO$_4$~\cite{https://doi.org/10.1111/j.1151-2916.2003.tb03575.x,2019Structural}. The continuous shift of the diffraction peak and monotonous increase in the lattice parameters $b$ and $c$ both indicate the homogeneous substitution of Co$^{2+}$ with Zn$^{2+}$ in \nczso.

\subsection{Magnetic susceptibility}

\begin{figure*}[htb]
\centerline{\includegraphics[width=6.8in]{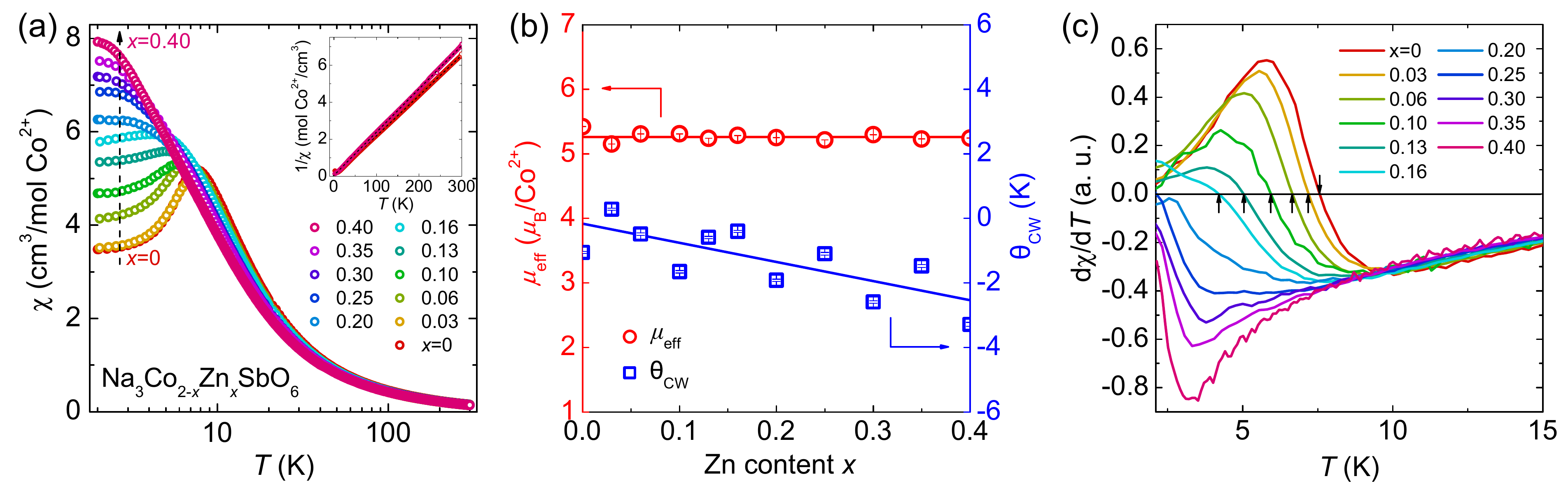}}
\caption{
 (a) Temperature dependence of dc magnetic susceptibility~($\chi$) of \nczso with 0~$\leq x\leq$~0.40. The measurements were performed in zero-field-cooling~(ZFC) condition. The inset shows the inverse susceptibility data of two representative zinc concentrations, which are $x$ = 0 and 0.35, respectively. The dashed lines are the fits with the Curie-Weiss law. (b) The extracted effective moment $\mu_{\rm eff}$ and Curie-Weiss temperature $\Theta_{\rm CW}$ as a function of $x$. The solid lines are guides to the eye. (c) The derivative of dc magnetic susceptibility versus temperature for different $x$. The arrows mark the positions of d$\chi$/d$T$ = 0 from which the AFM transition temperature $T_{\rm N}$ is extracted.
\label{fig2}}
\end{figure*}

Figure~\ref{fig2}(a) shows the magnetic properties of \nczso by measuring dc magnetic susceptibility~($\chi$) with a field of 0.1~T. For $x$ = 0 compound, there is a sharp maximum around 7.5~K with decreasing temperature, indicating an onset of the antiferromagnetic~(AFM) state. This result agrees with previous reports~\cite{WONG201618,PhysRevX.12.041024}. When introducing Zn$^{2+}$ into the system, the sharp maximum shifts to lower temperatures, and meanwhile it becomes much broader until indiscernible at $x\sim$~0.2. It indicates an efficient suppression of the AFM transition temperature by spin vacancies resulting from Zn$^{2+}$ occupying Co$^{2+}$ sites. We also note an obvious rise in the low-temperature magnetic susceptibility with increasing $x$. This behavior is also observed in Ir-diluted $\alpha$-RuCl$_3$, which is attributed to the uncompensated moments introduced by nonmangetic impurities in the magnetically ordered state~\cite{PhysRevLett.119.237203}. On the other hand, the susceptibility data above the AFM transition temperature are nearly overlapped for various $x$ and they follow the Curie-Weiss behavior well [see the inset in Fig.~\ref{fig2}(a)]. From a fit of $\chi$ within 50 and 300~K to $\chi_{\rm 0}$ + $C/(T-\Theta_{\rm CW}$), where $\chi_{\rm 0}$, $C$, $\Theta_{\rm CW}$ denote temperature-independent term related to nuclear and Van-Vleck paramagnetic contributions, Curie-Weiss constant, and Curie-Weiss temperature, respectively, we obtain $\chi_{\rm 0} \approx$ -8.78~$\times$~10$^{-3}$~cm$^3$/mol Co$^{2+}$, $C \approx$ 3.67~cm$^3$ K/mol Co$^{2+}$, and $\Theta_{\rm CW} \approx$ -1.06~K for $x$ = 0 compound. We also estimate the molecular diamagnetic susceptibility $\chi_{\rm D} \approx$ -1.30~$\times$~10$^{-4}$~cm$^3$/mol by summing diamagnetic Pascal's constants of different ions of \ncso, which is much less than $\chi_{\rm 0}$ and thus can be ignored. Then it yields an effective magnetic moment $\mu_{\rm eff}$ = 5.42$~\mu_{\rm B}$/Co$^{2+}$, where $\mu_{\rm B}$ is the Bohr magneton. This value is much larger than that of 3.87~$\mu_{\rm B}$ for a spin-only $S$ = 3/2 effective moment, reflecting a significant contribution from unquenched orbital angular momentum in \ncso~\cite{WONG201618,PhysRevMaterials.3.074405}. Following the way of processing susceptibility data of the $x$ = 0 compound, we also extract the values of $\mu_{\rm eff}$ and $\Theta_{\rm CW}$ for the $x$ $>$ 0 series, and the results are shown in Fig.~\ref{fig2}(b). The effective moment of Co$^{2+}$ remains almost constant for $0\leq x\leq0.40$. It implies that Co$^{2+}$ maintains such a spin-orbit entangled state throughout the Zn-doped process. In the meanwhile, we also note an unexpected increase in $|\Theta_{\rm CW}|$ that manifests an enhanced magnetic coupling. It is an unusual behavior since spin vacancies generally hinder the exchange paths and thus weaken its strength. We speculate that Zn doping should have a more complex impact on the magnetic ground states, not just diluting the magnetism of the system. To quantitatively obtain the AFM transition temperatures and reveal the evolution of magnetic ground states of \nczso more clearly, we present the differential susceptibility d$\chi$/d$T$ in Fig.~\ref{fig2}(c). The positions where d$\chi$/d$T$ = 0 are marked by the arrows, and they actually correspond to the maxima of the $\chi$-$T$ curves displayed in Fig.~\ref{fig2}(a). One can see it shifts to lower temperatures with increasing $x$ until vanishes when $x\geq$~0.2. This result indicates that $x$~=~0.2 is a critical point where the AFM order is completely suppressed.

\begin{figure}[h!]
  \centering
  \includegraphics[width=0.98\linewidth]{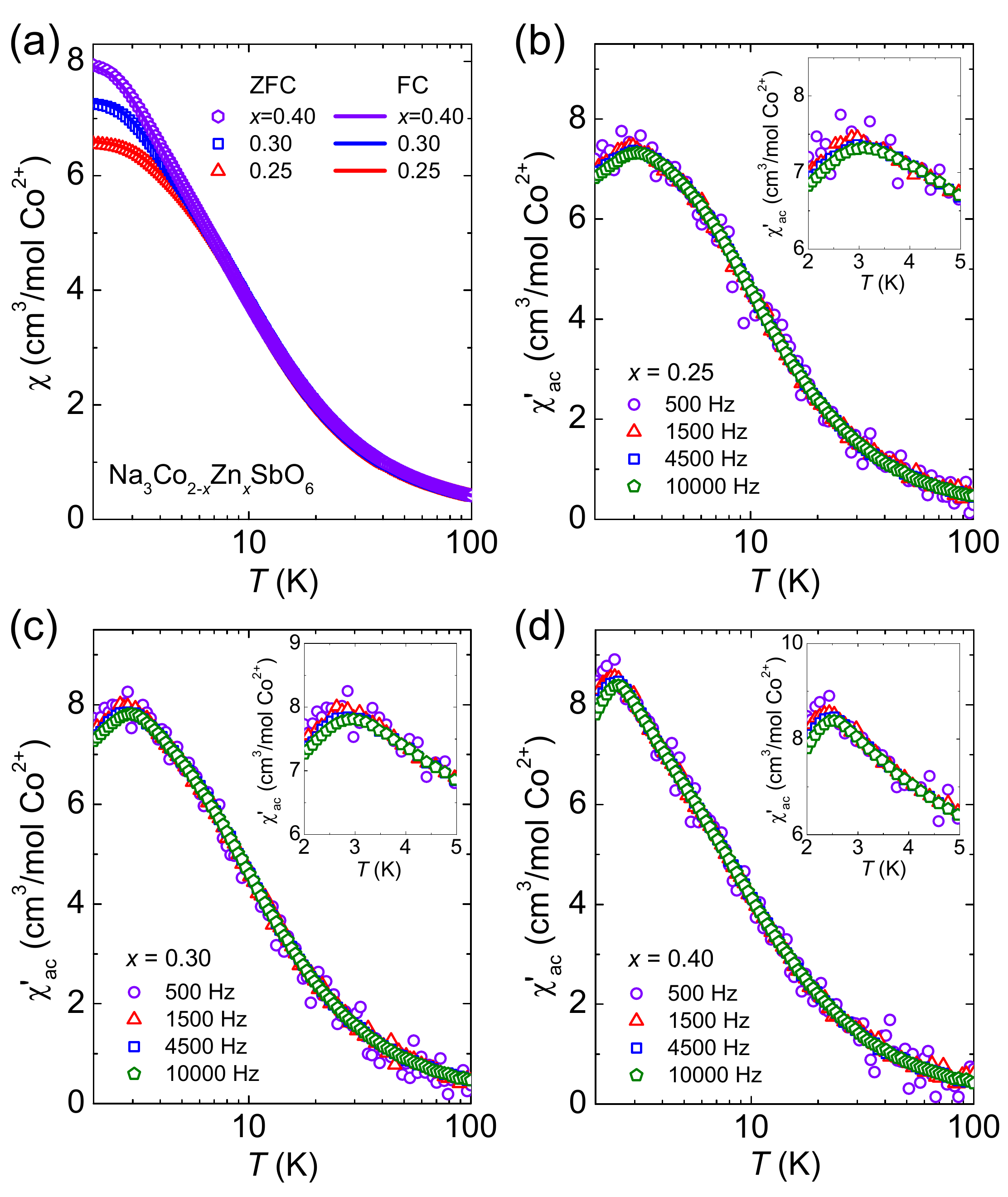}
  \caption{
  (a) Temperature dependence of dc magnetic susceptibility of several typical compounds with $x$ = 0.25, 0.30, and 0.40, measured in both ZFC and FC conditions. (b)-(d) Temperature dependence of the real part of ac magnetic susceptibility~($\chi^{\prime}_{\rm ac}$) with different driving frequencies for $x$ = 0.25, 0.30, and 0.40 compounds, respectively. The insets show the zoom-in view of the low-temperature data.}
  \label{fig3}
\end{figure}

Considering the fact that introducing Zn dopants will result in structural disorder due to site mixing of Zn$^{2+}$ and Co$^{2+}$, which is a main ingredient of a spin glass in the magnetically frustrated system~\cite{RevModPhys.58.801,Mydosh1993}, we specially measure dc magnetic susceptibility in both ZFC and field-cooling~(FC) conditions on several magnetically disordered compounds. As shown in Fig.~\ref{fig3}(a), there is no any signature of bifurcation between ZFC and FC curves, indicative of no spin freezing at low temperatures. Moreover, we also measure the ac magnetic susceptibility of these samples with several different driving frequencies. The results in Figs.~\ref{fig3}(b)-\ref{fig3}(d) show that there is no frequency dependence of the real part of ac magnetic susceptibility~($\chi^{\prime}_{\rm ac}$) despite a broad hump around 3~K. This behavior is also observed in another Co-based quantum spin liquid candidate Na$_2$BaCo(PO$_4$)$_2$ where such a hump occurs at lower temperature, and it is not attributed to a long-range phase transition or spin-glass transition~\cite{Zhong14505}. We believe it is true for the measured samples here. We speculate that the hump suggests a certain collective spin behavior that deserves further investigation in the future. Thus, the magnetically disordered regime may be a quantum spin liquid state with persistent dynamical spins that will be discussed in more detail next.

\subsection{Specific heat}

\begin{figure}[h!]
  \centering
  \includegraphics[width=0.98\linewidth]{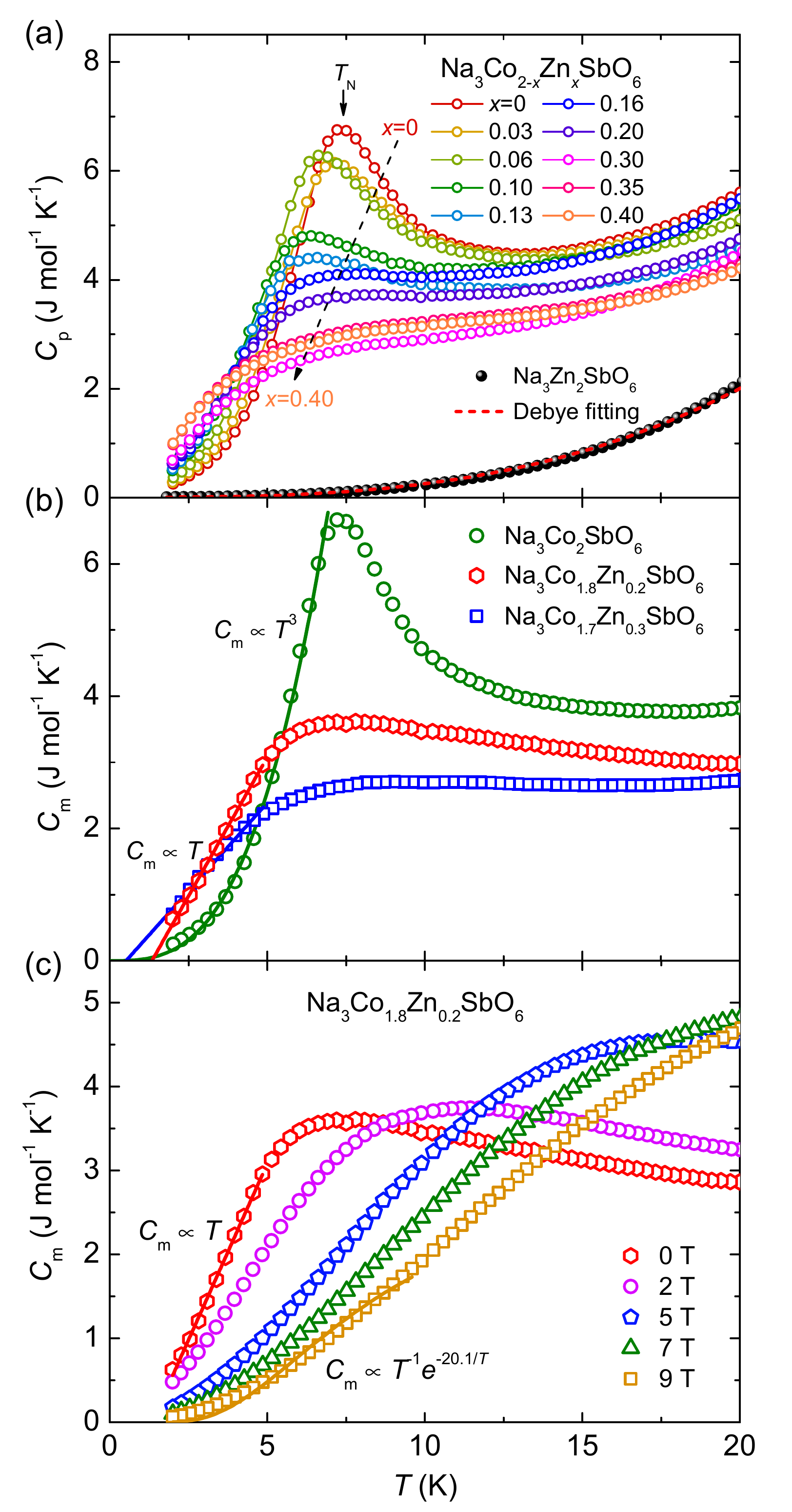}
  \caption{
  (a) Temperature dependence of specific heat~($C_{\rm p}$) of \nczso over a temperature range of 2-20~K at zero field. The dashed line with an arrow denotes an evolution tendency of the AFM phase transition with increasing $x$. Specific heat of a nonmagnetic reference compound Na$_3$Zn$_2$SbO$_6$ is also shown for comparison, which can be nicely fitted by a Debye model as $C_{\rm p}\propto T^3$. The dashed line denotes such a fitting. (b) Magnetic specific heat~($C_{\rm m}$) of the parent compound as well as the magnetically disordered compounds with two typical Zn content $x$ = 0.2 and 0.3. $C_{\rm m}$ is obtained by subtracting the lattice contribution using an isostructural nonmagnetic sample Na$_3$Zn$_2$SbO$_6$ by a scaling factor of 1.02. (c) Low-temperature magnetic specific heat of Na$_3$Co$_{1.8}$Zn$_{0.2}$SbO$_6$ compound measured at various magnetic fields. Solid lines in (b) and (c) are fits to the data described in the main text.}
  \label{fig4}
\end{figure}

Specific heat~($C_{\rm p}$) is a powerful technique to probe magnetic phase transitions and low-energy magnetic excitations~\cite{np4_459,nc2_275,RevModPhys.89.025003,npjqm4_12}. In order to further clarify the evolution process of the magnetic ground states of \nczso, we perform specific heat measurements on the compound series. In Fig.~\ref{fig4}(a), we depict the results of zero-field specific heat for $0\leq x\leq0.4$ compounds as well as an isostructural nonmagnetic compound Na$_3$Zn$_2$SbO$_6$. At $x$~=~0, there is a $\lambda$-type anomaly observed at 7.5~K, which signals an onset of long-range AFM state~\cite{WONG201618,PhysRevMaterials.3.074405}. The transition temperature is in good agreement with that determined in magnetic susceptibility. As $x$ increases, the anomaly peak shifts to lower temperatures and becomes rather broad and weak. Then it evolves into a mild hump when $x$ $\geq$~0.16. Such a hump in specific heat is no longer an indication of long-range magnetic phase transition, but a universal feature in spin glasses and QSL candidates, which may be associated with the establishment of short-range spin correlations~\cite{RevModPhys.58.801,Mydosh1993,prl98_107204,np4_459,nc2_275,sr5_16419,RevModPhys.89.025003,PhysRevLett.120.087201}. However, the aforementioned magnetic susceptibility measurements under ZFC and FC conditions indicate that a spin-glass state is unlikely. To further elucidate the nature of the magnetically disordered phase for $x$ $>$ 0.16, we attempt to extract the magnetic specific heat~($C_{\rm m}$) of $x$~=~0.2 and 0.3 compounds as well as the parent compound for comparison. In general, the total specific heat of a crystallized system with gapless elementary excitations can be represented as $C_{\rm p} = \gamma T + \beta T^3$~(Refs.~\onlinecite{np4_459,nc2_275}), where the temperature-linear and cubic terms denote fermionic and bosonic quasiparticle contributions, respectively. To extract the magnetic specific heat of \nczso more accurately, we perform a scaling procedure on their phononic specific heats with the total specific heat of an isostructural nonmangetic compound Na$_3$Zn$_2$SbO$_6$, as performed in $\alpha$-Ru$_{1-x}$Ir$_x$Cl$_3$~\cite{PhysRevB.99.214410}. There is a scaling factor of $\theta_{\rm D}$/$\theta_{\rm D,Na3Zn2SbO6}$ = 1.02. The so-obtained results are shown in Fig.~\ref{fig4}(b). For the parent compound, we clearly distinguish the AFM transition characterized by a $\lambda$-type anomaly around 7.5~K. Below it, the well-defined magnon excitations can be perfectly described by a cubic function $C_{\rm m} = \beta T^3$ with $\beta$ = 2.04(2)~$\times$~10$^{-2}$~J~mol$^{-1}$~K$^{-2}$. For the magnetically disordered compounds with $x$~=~0.2 and 0.3, a broad hump replaces the $\lambda$-type anomaly observed at $x$~=~0, which may indicate the establishment of short-range spin correlations as mentioned above. As temperature continues to decrease, magnetic specific heat drops steeply, and it can be nicely fitted by a linearly temperature-dependent term with $C_{\rm m}\propto T$. Then it gives rise to a significant Sommerfeld coefficient $\gamma$ = 0.84(1) and 0.53(2)~J~mol$^{-1}$~K$^{-2}$ over a fitting temperature range of 2-5~K for $x$ = 0.2 and 0.3 compounds, respectively. The behavior that a  temperature-cubic term $\beta T^3$ in magnetic specific heat develops into a temperature-linear one $\gamma T$ may suggest the evolution of magnetic ground states from well-defined magnon excitations with $S$ = 1 to itinerant quasiparticle excitations with $S$ = 1/2. The finite value of $\gamma$ in a magnetically disordered regime is quite unusual for an insulating system with no free electrons, and it is reminiscent of a gapless QSL state with fractional spinon excitations~\cite{prl98_107204,np4_459,nc2_275,PhysRevLett.122.167202}.

We further investigate the magnetic field dependence of magnetic specific heat with the  $x$~=~0.2 compound to track the behaviors of the low-energy excitations. In Fig.~\ref{fig4}(c), one can see that the hump is continuously suppressed until indistinguishable under a magnetic field of 9~T. This behavior is also observed in other QSL candidates, such as herbertsmithite~\cite{prl98_107204}, YbMgGaO$_4$~\cite{sr5_16419}, and YbZnGaO$_4$~\cite{PhysRevLett.120.087201}. The applied magnetic field is actually detrimental to the magnetically disordered state, since it may induce symmetry breaking and force the spin moments to align towards the field direction, and eventually drives the system into a ferromagnetic phase as long as the field is large enough~\cite{ZhenMa:106101,PhysRevB.104.224433}. The suppression of the hump and the decrease in low-temperature magnetic specific heat reflect that the magnetic ground state is continuously  meditated by an external field. In fact, magnetization of this regime easily becomes saturated under a moderate field of around 4~T~\cite{PhysRevMaterials.3.074405}. As a result, the power-law behavior gradually turns into an exponential one as the field increases up to 9~T, which is attributed to the evolution from gapless spinon-like excitations at zero field to gapped magnons in the fully polarized state. We fit the data with a gap function $C_{\rm m}\propto T^{\rm -1}{\rm exp}(-\Delta/T)$~(Ref.~\onlinecite{PhysRevB.107.054411}), where $\Delta$ denotes the spin gap. The fit yields a $\Delta$ value of 20.1(6)~K~($\sim$1.73~meV).

\subsection{Phase diagram}

\begin{figure}[htb]
  \centering
  \includegraphics[width=0.98\linewidth]{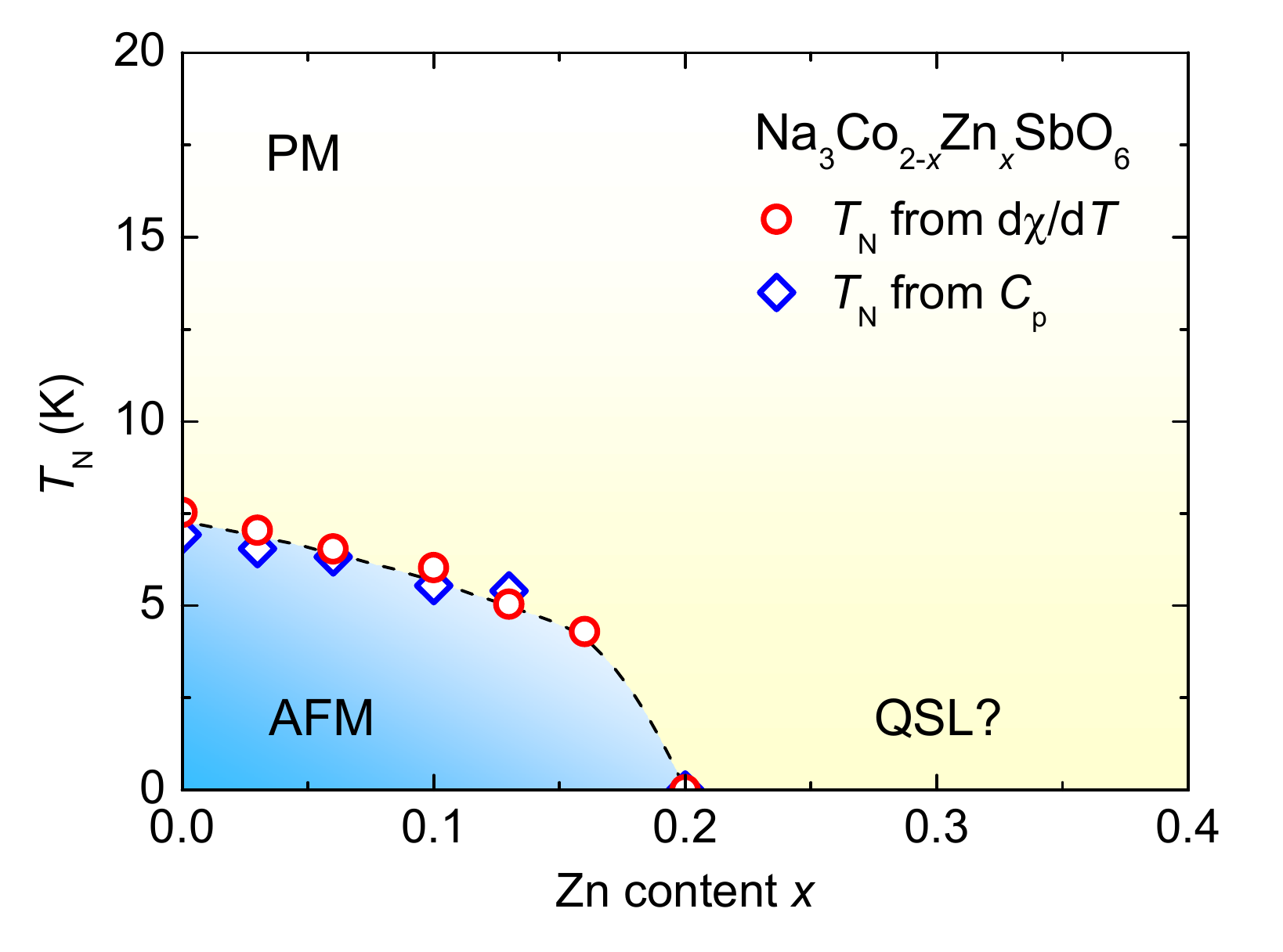}
  \caption{
  Magnetic phase diagram of \nczso. The higher white zone represents a trivial paramagnetic~(PM) state. The lower blue and yellow zones denote AFM and possible QSL states, respectively. The phase boundary denoted by a dashed line is determined from the differential susceptibility and specific heat data shown in Figs.~\ref{fig2}(c) and \ref{fig4}(a).}
  \label{fig5}
\end{figure}

By summarizing the aforementioned magnetic susceptibility and specific heat results, we finally establish a magnetic phase diagram of \nczso as shown in Fig.~\ref{fig5}. At $x$ = 0, the system undergoes an AFM phase transition at 7.5~K with decreasing temperature, and it enters the zigzag-type magnetic ground state~\cite{WONG201618,PhysRevMaterials.3.074405,PhysRevX.12.041024}. When incorporating nonmagnetic Zn$^{2+}$ into the regime, the transition temperature is continuously suppressed until it vanishes at $x$ = 0.2. The phase boundary between AFM and PM states are determined by extracting the zero value of the differential susceptibility d$\chi$/d$T$ as shown in Fig.~\ref{fig2}(c). It is also confirmed by the specific heat data featuring $\lambda$-type peaks as depicted in Fig.~\ref{fig4}(a). The results above suggest that long-range magnetic order of the parent compound is sensitive to spin vacancies, and a small amount of Zn doping with only 10\% can completely destroy it and drive the system into a magnetically disordered state. Furthermore, a finite linear term of magnetic specific heat in stark contrast to the $C_{\rm m}\propto T^3$ behavior of the parent compound as displayed in Fig.~\ref{fig4}(b) points to a possible gapless QSL state with fractional spin excitations in the magnetically disordered regime. Thus, the phase diagram of \nczso consists of three distinct zones, including a high-temperature PM phase, and low-temperature AFM and potential QSL states. Due to the lack of characteristic temperature to distinguish PM and QSL states in magnetic susceptibility and specific heat measurements, the boundary between them is indistinct.

\section{Discussions}

Since \ncso is a representative Kitaev magnet in 3$d$ transition-metal compounds, we think it is necessary to relate our results to the heavily studied 4$d$/5$d$ counterparts. It is found that zigzag orders in 4$d$ $\alpha$-RuCl$_3$~\cite{PhysRevLett.119.237203,PhysRevB.98.014407,PhysRevLett.124.047204} and 5$d$ Na$_2$IrO$_3$~\cite{PhysRevB.89.241102} are very sensitive to spin vacancies. Lightly doped $\alpha$-Ru$_{1-x}$Ir$_x$Cl$_3$ with $x$ = 0.2 and Na$_2$Ir$_{1-x}$Ti$_x$O$_3$ with $x$ = 0.05 have already turned them into a possible QSL~\cite{PhysRevLett.119.237203} and a spin glass~\cite{PhysRevB.89.241102}, respectively. Compared with them, the case in 3$d$ \nczso seems more like that of the former, despite that a lower Zn-doping level with only 10\% destroys the zigzag magnetic order and then induces a magnetically disordered phase. The fact that in \ncso the magnetic order is more easily suppressed with doping indicates that \ncso is likely to be more proximate to a pure Kitaev QSL phase than $\alpha$-RuCl$_3$. For the magnetically disordered phase in \nczso, there is no spin freezing observed either from our magnetic susceptibility measurements, indicating the persistent dynamical fluctuations down to 2~K. More importantly, the presence of a residual linear term in specific heat at zero field evidences a gapless QSL state with fractional quasipartical excitations~\cite{RevModPhys.89.025003,Broholmeaay0668,prl98_107204,np4_459,nc2_275,PhysRevLett.122.167202}.  Such a linearly temperature-dependent term with $\gamma$ = 0.020 and 0.0199~J~mol$^{-1}$~K$^{-2}$ was also captured in well studied organic salts $\kappa$-(BEDT-TTF)$_2$Cu$_2$(CN)$_3$ and EtMe$_3$Sb[Pd(dmit)$_2]_2$, respectively. In \nczso, we find $\gamma$ = 0.84(1) and 0.53(2)~J~mol$^{-1}$~K$^{-2}$ for $x$ = 0.2 and 0.3 compounds, respectively.  Since $\gamma$ is considered to be proportional to be the spinon density of states~\cite{np4_459}, such a large $\gamma$ value compared with those in $\kappa$-(BEDT-TTF)$_2$Cu$_2$(CN)$_3$ and EtMe$_3$Sb[Pd(dmit)$_2]_2$ may reflect much larger density of states at low energies in \nczso. Although it is in stark contrast with a quasi-quadratic behavior of magnetic specific heat for $\alpha$-Ru$_{0.8}$Ir$_{0.2}$Cl$_3$ suggestive of gapless Dirac-like excitations~\cite{PhysRevLett.124.047204}, they both exhibit positive signals of fermionic excitations expected for a QSL~\cite{np4_459,PhysRevLett.98.117205}. In other words, diluted Kitaev magnets may serve as a potential platform to host a variety of QSL phases as reported by recent both theoretical and experimental works~\cite{PhysRevLett.122.167202,PhysRevLett.124.047204,PhysRevX.11.011034,PhysRevLett.129.037204}. Moreover, we have also performed such a Zn-doping inverstigation on another honeycomb-lattice Kitaev magnet Na$_2$Co$_2$TeO$_6$. The results show that the AFM ground state of the parent compound can survive with Zn content $x$ up to $\sim$1.0~(Ref.~\onlinecite{PhysRevMaterials.7.014407}). Then it evolves into a spin-glass phase with short-range order. Compared with the results of \nczso in this paper, the insensitive response to the nonmagnetic Zn doping in Na$_2$Co$_{2-x}$Zn$_x$TeO$_6$ may signify magnetic couplings beyond the nearest-neighbor interactions in the parent compound, such as a leading third-nearest-neighbor Heisenberg interaction $J_{\rm 3}$ determined by INS measurements on single crystals~\cite{PhysRevLett.129.147202}. That means the destruction of longer-range magnetic couplings needs a higher-doping level of nonmagnetic ions. Then it will restrict the spin correlations in a short-range zone like spin clusters and meanwhile introduce serious structural disorder effect with increasing $x$ up to $\sim$1.0, which results in the formation of a spin-glass state~\cite{PhysRevMaterials.7.014407}. By contrast, the spin Hamiltonian of \ncso mainly contains the nearest-neighbor terms~\cite{PhysRevB.102.224429,PhysRevB.106.014413}. In this case, a small amount of spin vacancies generated by nonmagnetic doping can play an effective role in eliminating the perturbation of non-Kitaev interactions like the case of $\alpha$-Ru$_{1-x}$Ir$_x$Cl$_3$~(Ref.~\onlinecite{PhysRevLett.124.047204}). Additionally, there are recent reports arguing that Na$_2$Co$_2$TeO$_6$ harbors a triple-Q magnetic structure rather than a traditional zigzag order that has been determined in \ncso~(Refs.~\onlinecite{PhysRevB.103.L180404,PhysRevLett.129.147202}). It reflects that these two cobaltates host slightly different underlying physics even though they have the same Co-O layers since their crystal structure and space group are different~\cite{PhysRevMaterials.3.074405,PhysRevMaterials.7.014407}.

Given that Zn doping will inevitably introduce site and bond disorder into the system, which can contribute to the low-energy density of states and mimic the QSL features~\cite{PhysRevLett.122.167202,PhysRevB.104.035116,PhysRevX.11.011034,PhysRevLett.129.037204}, it calls for further theoretical calculations and more experimental verifications at lower temperatures, such as INS to measure the magnetic spectra and extract magnetic interactions, muon-spin relaxation to check the dynamical moments, and thermal conductivity to probe the itinerant quasiparticles. In particular, the investigations upon single crystals will bring more comprehensive information. Anyway, the introduction of nonmagnetic Zn$^{2+}$ in the 3$d$ Kitaev magnet \ncso give rise to a quite fascinating phase diagram that includes a probable QSL phase in the lightly doped regime.

\section{Summary}

To summarize, we have performed XRD, magnetic susceptibility, and specific heat characterizations to investigate the magnetic ground states of a Zn-substituted Kitaev magnet \ncso. Due to the quite close radii of Zn$^{2+}$ and high-spin Co$^{2+}$ ions, there is no structural transition but tiny changes on the lattice parameters over a substitution range of $0\leq x\leq0.4$. Magnetic susceptibility and specific heat measurements both show that AFM order of the parent compound is sensitive to Zn doping, and a low doping level with $x$~=~0.2 can completely destroy it and drive the system into a magnetically disordered state. More importantly, the presence of a residual linear term in the specific heat below 5~K evidences a possible gapless QSL state with itinerant fermionic excitations in the $x \geq 0.2$ regime. These results demonstrate that the introduction of spin vacancies by chemical substitution may serve as a feasible means to achieve the intrinsic QSL state in Kitaev magnets.

\section{Acknowledgments}

We thank Z.-Y. Dong for helpful discussions. The work was supported by the National Key Projects for Research and Development of China with Grant No.~2021YFA1400400, National Natural Science Foundation of China with Grants No.~12204160, No.~12225407, No.~12074174, No.~12074111, No.~12061130200, No.~11961160699, No.~11974392, and No.~12274444, Hubei Provincial Natural Science Foundation of China with Grant No.~2021CFB238, China Postdoctoral Science Foundation with Grants No.~2022M711569 and No.~2022T150315, Jiangsu Province Excellent Postdoctoral Program with Grant No.~20220ZB5, K. C. Wong Education Foundation with Grant No.~GJTD-2020-01, the Youth Innovation Promotion Association of the CAS with Grant No.~Y202001, Beijing Natural Science Foundation with Grant No.~JQ19002, and Fundamental Research Funds for the Central Universities. Z.M. thanks Beijing National Laboratory for Condensed Matter Physics for funding support.

%

\end{document}